\documentclass[10pt,journal,compsoc]{IEEEtran}

\usepackage{amsmath,amsfonts}
\usepackage{graphicx}
\usepackage{cite}
\usepackage{url}
\usepackage[ruled,vlined]{algorithm2e}
\usepackage{hyperref}
\usepackage{xcolor}
\usepackage{booktabs} 
\usepackage{comment}

\begin{document}

\title{Word2VecGD: Neural Graph Drawing with Cosine-Stress Optimization}

\author{Minglai Yang, Reyan Ahmed}

\IEEEtitleabstractindextext{%
\begin{abstract}
We propose a novel graph visualization method leveraging random walk-based embeddings to replace costly graph-theoretical distance computations. Using word2vec-inspired embeddings, our approach captures both structural and semantic relationships efficiently. Instead of relying on exact shortest-path distances, we optimize layouts using cosine dissimilarities, significantly reducing computational overhead. Our framework integrates differentiable stress optimization with stochastic gradient descent (SGD), supporting multi-criteria layout objectives. Experimental results demonstrate that our method produces high-quality, semantically meaningful layouts while efficiently scaling to large graphs. Code available at: \url{https://github.com/mlyann/graphv_nn}
\end{abstract}

\begin{IEEEkeywords}
Graph Drawing, Graph Neural Networks, Embeddings, Gradient Descent, Multi-Criteria Optimization
\end{IEEEkeywords}}

\maketitle

\section{Introduction}

Visualization of complex graphs is a longstanding challenge in fields such as network science, biology, and information retrieval. High-quality visual layouts are essential to identifying latent structures, communities, and anomalies at a glance. Yet, classical graph drawing methods—ranging from stress minimization to force-directed algorithms—rely heavily on shortest-path computations or intricate heuristic forces. These dependencies make them computationally expensive and limit their scalability to large graphs.

Recent advances in representation learning have opened new opportunities for graph visualization. Methods such as DeepWalk~\cite{perozzi2014deepwalk}, Node2Vec~\cite{grover2016node2vec}, and graph neural networks (GNNs)~\cite{kipf2016semi, velivckovic2017graph} learn low-dimensional node embeddings that capture both local and global structure. These embeddings have proven instrumental in downstream tasks like node classification and link prediction, but they have not been fully leveraged for scalable, high-quality visual layouts.

In this paper, we introduce \emph{Word2VecGD}, a framework that integrates representation learning, differentiable optimization, and multi-criteria layout objectives into a unified pipeline. Instead of relying on shortest-path distances, we adopt cosine dissimilarities derived from embeddings trained via random walks and word2vec-like objectives. By replacing costly distance calculations with embedding-based metrics, we efficiently scale to large graphs without sacrificing semantic coherence.

Our method encodes node positions as parameters in a fully differentiable objective, supporting multiple criteria (e.g., stress, neighborhood preservation, and edge length uniformity) via additive loss functions. This contrasts with traditional layout techniques, which require custom force models or complex heuristics for each criterion. By formulating graph drawing as a differentiable optimization problem solvable by standard stochastic gradient descent (SGD), we enable flexible, plug-and-play integration of new readability metrics.

Empirical evaluations on synthetic and real-world graphs demonstrate that Word2VecGD achieves comparable or better quality than classical methods, while scaling more gracefully and producing semantically meaningful layouts. Our approach bridges the gap between the efficiency and interpretability of learned embeddings and the expressiveness of multi-criteria graph visualization, thus offering a novel machine learning-driven paradigm for large-scale graph drawing.

\section{Related Work}

\subsection{Classical and Approximate Graph Drawing}
Classical graph layout algorithms aim to produce visually coherent representations by optimizing stress or applying force-directed heuristics. Kamada and Kawai~\cite{kamada1989algorithm} proposed stress minimization, aligning Euclidean distances in the layout with graph-theoretic shortest-path distances. Although effective for small graphs, this approach is computationally expensive due to the need for all-pairs shortest-path computations. Force-directed methods~\cite{fruchterman1991graph} use attractive and repulsive forces to iteratively refine layouts, balancing computational cost with layout quality. However, these methods often struggle to scale to large networks.

Approximation techniques, such as PivotMDS~\cite{brandes2004solving} and scalable force-directed placement (sfdp)~\cite{ellson2001graphviz}, reduce computational overhead through sampling or hierarchical decomposition. Despite these advancements, such methods still rely on graph-theoretic distances, limiting their scalability. 
Many of these algorithms require the all-pairs shortest paths, which can be computationally expensive~\cite{cormen2022introduction}.

\subsection{Representation Learning for Graphs}
Graph embedding techniques redefine network analysis and visualization by mapping nodes into continuous vector spaces that preserve structural and semantic relationships. DeepWalk~\cite{perozzi2014deepwalk} and Node2Vec~\cite{grover2016node2vec} employ random walks and word2vec-style objectives to generate embeddings encoding both local and global graph structures. Graph neural networks (GNNs)~\cite{kipf2016semi,velivckovic2017graph} further extend these methods by incorporating node features and task-specific message passing.

While embedding-based methods have become standard for predictive and inference tasks, their integration into graph visualization is still underexplored. Current approaches often rely on static embeddings, limiting their adaptability to dynamic or multi-criteria layout objectives and their scalability to very large networks.

\subsection{Multi-Criteria Graph Drawing}
Ahmed et al.~\cite{ahmed2020graphdrawinggradientdescent} introduced (GD)\(^2\), a framework combining embeddings with gradient descent for layout optimization. 
Ahmed et al. extended this with (SGD)\(^2\)~\cite{ahmed2021multicriteriascalablegraphdrawing}, which incorporates multiple readability criteria into the layout process using stochastic gradient descent, enabling scalability to graphs with millions of nodes. Scalable tree layouts~\cite{ahmed2023tree} address specific visualization challenges.


\subsection{Differentiable Optimization for Graph Layouts}
Recent advancements have shown the potential of differentiable frameworks for graph layout optimization. Ahmed et al.~\cite{ahmed2020graphdrawinggradientdescent,ahmed2021multicriteriascalablegraphdrawing} employed smooth, differentiable loss functions to incorporate multiple layout objectives, optimizing these with stochastic gradient descent. By treating node positions as learnable parameters, these methods achieve scalable and flexible layouts.

Building on this foundation, our approach replaces shortest-path-based stress with cosine dissimilarities derived from word2vec-style embeddings. By integrating multiple readability criteria directly into the optimization objective, our method supports scalable and dynamic refinement for large and semantically complex graphs, further extending the applicability of embedding-driven visualization frameworks.

\section{Methods}
\subsection{Embedding-Based Stress Minimization}
\label{sec:sns}
Our method begins by deriving node embeddings from random walks on the input graph, treating nodes as ``words'' and random walks as ``sentences.'' We then train a word2vec-style skip-gram model to learn vector representations that capture both local and global structural properties. Using these embeddings, we compute cosine-based dissimilarities between nodes, replacing traditional graph-theoretic distances. Finally, we optimize a stress function using stochastic gradient descent (SGD) to produce high-quality graph layouts.

\noindent\textbf{Random Walks as Sentences.}
Given a graph $G=(V,E)$, we simulate random walks to generate ``node sequences'' analogous to sentences in a corpus:
\[
\text{walk}(v_0) = (v_0, v_1, \ldots, v_L),
\]
where $v_0$ is the start node, $L$ is the walk length, and each subsequent node $v_{i+1}$ is chosen uniformly at random from the neighbors of $v_i$. By repeating this process for multiple start nodes, we build a collection of walks:
\[
\mathcal{W} = \{\text{walk}(v) \mid v \in V\}.
\]

\noindent\textbf{Word2Vec-Style Embeddings.}
Treating each node as a ``word'' and each random walk as a ``sentence,'' we train a skip-gram word2vec model to learn node embeddings. For a center node $w$ and a context node $u$ within a fixed window size, the skip-gram model aims to maximize:
\[
\max_{\theta} \sum_{w \in \mathcal{W}} \sum_{u \in \mathcal{C}(w)} \log P(u \mid w;\theta),
\]
where $\mathcal{C}(w)$ is the set of context nodes for $w$, and $P(u \mid w;\theta)$ is given by the softmax:
\[
P(u \mid w;\theta) = \frac{\exp(\mathbf{v}_u \cdot \mathbf{v}_w)}{\sum_{u' \in V}\exp(\mathbf{v}_{u'} \cdot \mathbf{v}_w)}.
\]
Here, $\mathbf{v}_w \in \mathbb{R}^d$ is the embedding vector for node $w$, and $\theta$ represents all embedding parameters. Training this objective (e.g., using negative sampling for efficiency) yields embeddings that reflect node co-occurrence patterns from the random walks.

\noindent\textbf{Cosine Dissimilarity and Stress Minimization.}
Once the node embeddings $\{\mathbf{v}_i\}_{i=1}^{n}$ are learned, we define the cosine dissimilarity for each pair $(i, j)$ as:
\[
d_{\text{cos}}(i,j) = 1 - \frac{\mathbf{v}_i \cdot \mathbf{v}_j}{\|\mathbf{v}_i\|\|\mathbf{v}_j\|}.
\]

We then embed the graph into 2D by placing each node $i$ at position $X_i \in \mathbb{R}^2$. Instead of using shortest-path distances, we align Euclidean distances with $d_{\text{cos}}(i,j)$. The stress function is:
\[
L_{\text{CD}} = \sum_{i<j} w_{ij}\bigl(\|X_i - X_j\|_2 - d_{\text{cos}}(i,j)\bigr)^2,
\]
where $w_{ij} = d_{\text{cos}}(i,j)^{-2}$ normalizes the influence of each pair.

\noindent\textbf{Multi-Criteria Optimization via SGD.}
Additional readability criteria (e.g., ideal edge lengths, neighborhood preservation) can be incorporated by adding corresponding terms to the loss:
\[
L(X) = \sum_c \alpha_c L_c(X).
\]
Each $L_c(X)$ is chosen to be differentiable. We use stochastic gradient descent to update node positions:
\[
X \leftarrow X - \eta \nabla_X L(X),
\]
sampling subsets of edges, node pairs, or other elements at each iteration to ensure scalability. This approach efficiently handles large graphs, leverages embedding-based semantic relationships, and flexibly integrates multiple layout criteria.

\subsection{Scale-Normalized Stress (SNS)}

Scale-Normalized Stress (SNS)~\cite{ahmed2024sizematterscaleinvariantstress} eliminates scale sensitivity in graph layout evaluation, ensuring fair comparisons across layouts by incorporating an optimal scaling factor. Traditional metrics like Raw Stress (RS) and Normalized Stress (NS) are influenced by scale, leading to misleading results. SNS addresses this by minimizing stress with respect to layout scale.

The optimal scaling factor $\alpha_{\text{min}}$ is computed as:
\[
\alpha_{\text{min}} = \frac{\sum_{i<j} d_{ij}^{-1} ||X_i - X_j||}{\sum_{i<j} d_{ij}^{-2} ||X_i - X_j||^2},
\]
where $||X_i - X_j||$ is the Euclidean distance, and $d_{ij}$ is the graph-theoretic distance.

Using $\alpha_{\text{min}}$, SNS is defined as:
\[
\text{SNS} = \sum_{i<j} d_{ij}^{-2} \left( \alpha_{\text{min}} ||X_i - X_j|| - d_{ij} \right)^2.
\]

SNS ensures that evaluations focus solely on the fidelity of graph-theoretic distances, independent of layout size, making it a robust and fair metric for comparing algorithms.

\section{Experiment}
We first run a simple experiment where we don't use any model to learn the layouts. In this experiment, we run the algorithm described in Section~\ref{sec:sns}. We refer to our method as Word2Vec.
\subsection{Dataset}
We evaluate our method on a synthetic dataset of 4000 Erdo-Rényi graphs with varying sizes and densities. Graphs are generated with node sizes $n \in \{20, 30\}$ and edge probabilities $p \in \{0.2, 0.4, 0.6, 0.8\}$, with 50 instances per combination to ensure diversity. Each graph is represented by 2D embeddings learned via a Word2Vec model trained on random walks, capturing structural relationships. Stress minimization further refines node positions, ensuring alignment between embedding distances and graph topology. This dataset provides a controlled yet diverse benchmark for evaluating graph visualization techniques.

\subsection{Graph Visualization with Graphviz Layouts}

We visualize the synthetic graphs using various Graphviz layout algorithms~\cite{gansner2009drawing} to benchmark their performance. Each layout method offers a unique approach to arranging nodes based on the graph structure:

\begin{itemize}
    \item \textbf{Dot}: Hierarchical layouts for directed acyclic graphs.
    \item \textbf{Neato}: Force-directed layouts for general undirected graphs.
    \item \textbf{Fdp}: Force-directed layouts for medium graphs.
    \item \textbf{Sfdp}: Force-directed multiscale layouts for large graphs.
    \item \textbf{Twopi}: A radial layout ideal for hierarchical structures.
    \item \textbf{Circo}: Circular layout designed for cyclic graphs.
    \item \textbf{Patchwork}: A layout for modular clusters.
    \item \textbf{Osage}: A layout specialized for clustered graphs.
\end{itemize}

Figure~\ref{fig:all_methods} illustrates the results for each layout method. Each grid contains the same set of synthetic graphs, enabling direct comparison of node arrangements across different layout strategies.

\subsection{Results and Comparison of Layout Methods}

\begin{figure*}[ht]
    \centering
    \includegraphics[width=\linewidth]{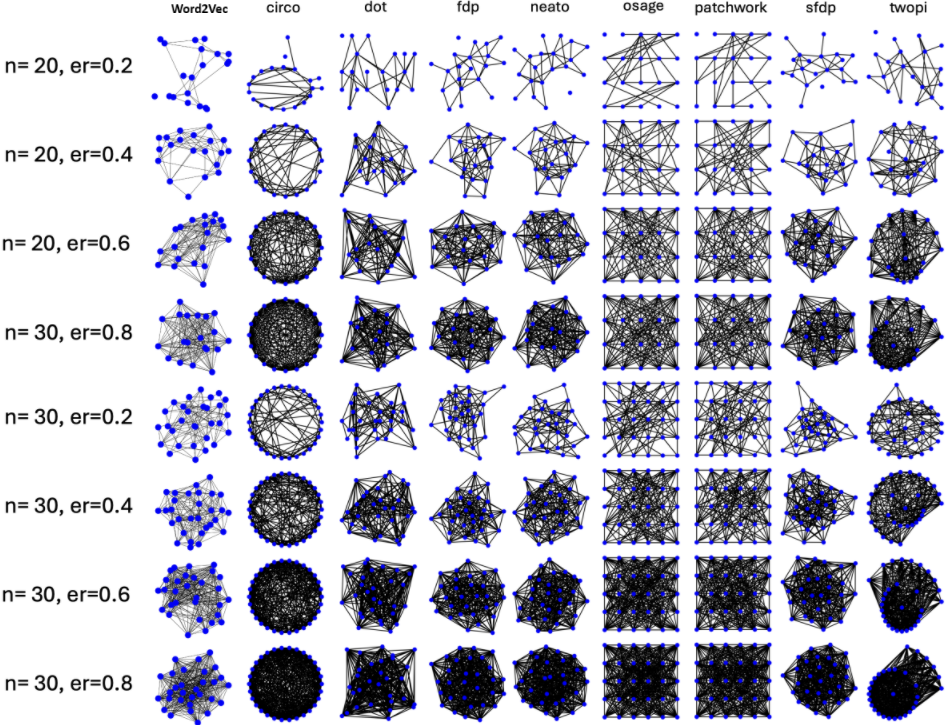} 
    \caption{Comparison of stress across all layout methods. 
    }
    \label{fig:all_methods}
\end{figure*}

We evaluated the performance of various layout methods on synthetic graphs with varying node sizes (\(n=20, 30\)) and edge probabilities (\(p=0.2, 0.4, 0.6, 0.8\)). The results, summarized in Table~\ref{tab:layout_comparison} and visualized in Figure~\ref{fig:all_methods}, provide insights into the capabilities and limitations of each method. Stress, the primary metric in this evaluation, quantifies how well layouts preserve graph structure, with lower values indicating better performance.
We can see that Neato, Fdf, and Sfdp are the best methods in terms of stress scores in Table~\ref{tab:layout_comparison}. Although Word2Vec is not achieving that well, in the following sections we will see that we can train a model that is more generalizable for larger datasets and faster than other models.

\begin{table*}[ht]
\centering
\caption{Average Stress Comparison Across Layout Methods. 
}
\label{tab:layout_comparison}
\begin{tabular}{c c c c c c c c c c}
\hline
\textbf{Node} & \textbf{p} & \textbf{Dot} & \textbf{Neato} & \textbf{Fdp} & \textbf{Sfdp} & \textbf{Twopi} & \textbf{Circo} & \textbf{Word2Vec} & \textbf{DeepGD} \\ \hline
20 & 0.2 & 22.23 & 8.98 & 9.98 & 9.44 & 24.44 & 21.42 & 18.96 & 15.96 \\ 
20 & 0.4 & 21.52 & 11.56 & 11.57 & 11.96 & 23.60 & 20.37 & 17.27 & 26.15 \\ 
20 & 0.6 & 21.11 & 12.72 & 12.81 & 12.86 & 21.48 & 19.43 & 22.17 & 29.04 \\ 
20 & 0.8 & 18.74 & 13.41 & 12.72 & 13.11 & 18.93 & 17.49 & 27.24 & 33.22 \\ 
30 & 0.2 & 56.47 & 26.05 & 26.76 & 27.63 & 56.21 & 50.59 & 57.05 & 38.08 \\ 
30 & 0.4 & 56.75 & 32.45 & 32.49 & 32.85 & 54.29 & 49.39 & 50.87 & 54.60 \\ 
30 & 0.6 & 56.82 & 34.92 & 34.54 & 35.06 & 52.43 & 47.93 & 47.47 & 79.63 \\ 
30 & 0.8 & 51.68 & 34.60 & 33.64 & 34.19 & 47.41 & 43.00 & 52.79 & 87.29 \\ \hline
\end{tabular}
\end{table*}

\subsection{Out\,-\,of\,-\,Distribution (OOD) Evaluation}
\label{sec:ood}

Real-world deployment rarely guarantees that a model will see graphs drawn
from the same size or density distribution it was trained on.
To probe the robustness of \textsc{Word2VecGD}, we conduct two
OOD studies—varying \emph{node cardinality} and \emph{edge density}—and
compare against \textsc{Neato} and \textsc{SFDP}.

\subsubsection{Datasets}

\textbf{Mixed--nodes.} Four stress-optimizer networks were trained on
graphs of \(\{18,19,21,22\}\) nodes (uniform ER probability~\(p=0.5\)),
with early-stopping every \(1000\) steps to avoid over-fitting.
At test-time, each model drew unseen \(20\)-, \(50\)-, and \(100\)-node
graphs.  For each size we generated \(100\) random instances and report the
mean (\(\mu\)) and standard deviation (\(\sigma\)) of
\emph{scale-normalized stress} (\textsc{sns}, cf.\ Section~\ref{sec:sns})
as well as wall-clock running times.

\textbf{Mixed--edges.} A second family of networks was trained on
\(20\)-node graphs whose ER edge probability
spanned \(\{0.1,0.2,0.3,0.4,0.6,0.7,0.8,0.9\}\).
All networks were evaluated on \(20\), \(50\), and \(100\)-node graphs
with the canonical density \(p=0.5\).



\subsubsection{Layout methods}

\textbf{Stress-optimizer neural network.} We adopt a residual MLP with \(\ell\!=\!80\) hidden layers of width 64,
batch-norm and 0.3 dropout after each block.
Optimization uses Adam (\(lr=5\!\times\!10^{-5}\),
\(\beta_1=0.9,\beta_2=0.999\)) for 10 000 steps, mini-batching
128 graphs per step.  Early stopping monitors scale-normalized stress
on a held-out validation split every 1000 steps.


\textbf{Baselines.}
We compare our method with Neato and Sfdp.
Both algorithms include their own multilevel initializations and thus
incur higher CPU runtime; no GPU acceleration is available.

\textbf{Metrics.} We report \emph{scale-normalized stress} (\textsc{sns}) as described
in Eq. (9), the gold-standard for structural fidelity that is immune
to layout scale.  
Runtime is decomposed into (i) embedding lookup (\textbf{EmbedT}),
(ii) forward pass (\textbf{FwdT}), and
(iii) end-to-end latency (\textbf{TotalT}).
For Neato and Sfdp a single wall-clock figure is recorded.



\subsubsection{Results}
\label{sec:results}

Table~\ref{tab:ood} summarizes the findings. Although our method (Word2VecGD) has higher stress scores, it is faster than the baselines. Additionally, notice that the model is generalizable since the testing dataset contains larger instances than the training dataset.

\begin{table*}[t]
\centering
\caption{Out-of-distribution performance and runtime.  Mean $\pm$ std over 100 random graphs per row.  All times in seconds.}
\label{tab:ood}
\resizebox{\textwidth}{!}{%
\begin{tabular}{@{}l c c c c c c c c c c c c@{}}
\toprule
\textbf{Setting} & \textbf{$n$} &
\multicolumn{5}{c}{\textbf{Neural (\textsc{Word2VecGD})}} &
\multicolumn{3}{c}{\textbf{Neato}} &
\multicolumn{3}{c}{\textbf{SFDP}}\\
\cmidrule(lr){3-7}\cmidrule(lr){8-10}\cmidrule(l){11-13}
 & &
$\mu_{\textsc{sns}}\!\downarrow$ & $\sigma_{\textsc{sns}}$ &
EmbedT$\!\downarrow$ & FwdT$\!\downarrow$ & TotalT$\!\downarrow$ &
$\mu_{\textsc{sns}}$ & $\sigma_{\textsc{sns}}$ & Time$\!\downarrow$ &
$\mu_{\textsc{sns}}$ & $\sigma_{\textsc{sns}}$ & Time$\!\downarrow$\\
\midrule
mixed--nodes & 20  & 48.83 & 2.18 & $4.9\!\times\!10^{-5}$ & 0.0068 & 0.0069 & 24.81 & 1.90 & 0.219 & 25.37 & 1.77 & 0.217\\
             & 50  & 345.78 & 6.73 & $1.2\!\times\!10^{-4}$ & 0.0073 & 0.0075 & 231.64 & 5.30 & 1.081 & 234.45 & 4.88 & 1.087\\
             & 100 & 1756.04 & 14.56 & $2.1\!\times\!10^{-4}$ & 0.0068 & 0.0070 & 1070.19 & 11.00 & 4.063 & 1084.73 & 10.03 & 4.059\\[2pt]
mixed--edges & 20  & 50.42 & 2.39 & $5.0\!\times\!10^{-5}$ & 0.0068 & 0.0069 & 24.81 & 1.90 & 0.216 & 25.37 & 1.77 & 0.217\\
             & 50  & 385.77 & 7.96 & $1.3\!\times\!10^{-4}$ & 0.0080 & 0.0081 & 231.64 & 5.30 & 1.093 & 234.45 & 4.88 & 1.097\\
             & 100 & 1451.62 & 14.82 & $2.1\!\times\!10^{-4}$ & 0.0067 & 0.0069 & 1070.19 & 11.00 & 4.185 & 1084.73 & 10.03 & 4.183\\
\bottomrule
\end{tabular}}
\end{table*}

\section{Conclusion}

We have studied a method that captures the graph structure through random walks, and by utilizing this method, we can bypass the computationally expensive all-pairs shortest path distances. We then propose another method that builds upon the first and can produce layouts faster compared to some well-known layout methods. We primarily focused on stress, but it would be interesting to try other metrics and their combinations. Additionally, it would be interesting to run the model on larger datasets.

\bibliographystyle{IEEEtran}
\bibliography{bib}
\newpage
\section{Appendix}
This appendix provides visualizations of various graph layout strategies used in our evaluation. Each layout method is applied to the same set of synthetic graphs to facilitate side-by-side comparison.

\begin{figure*}[ht]
    \centering
    \includegraphics[width=\textwidth]{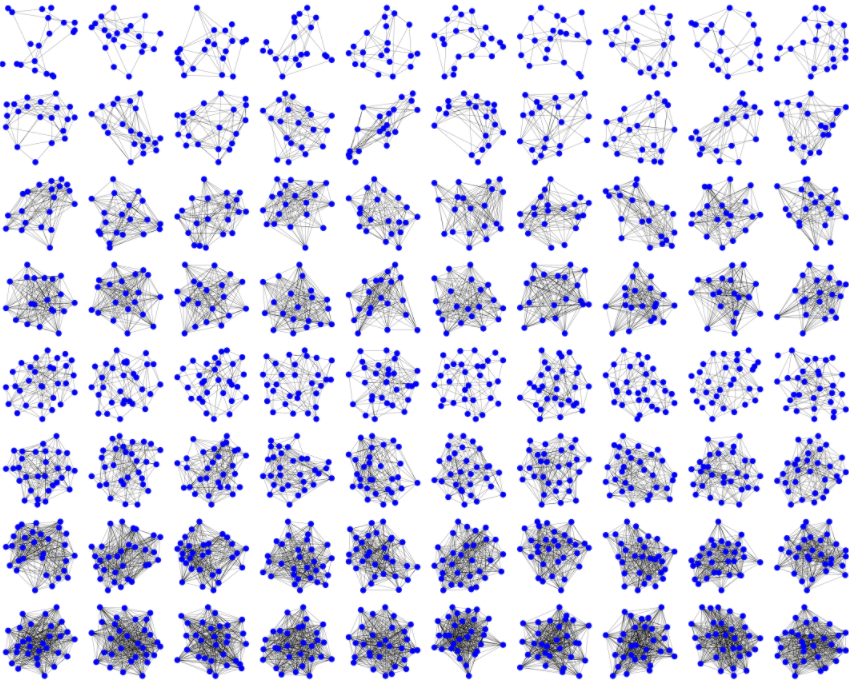}
    \caption{Visualization produced by our embedding-based method (\textsc{Word2VecGD}). Node positions are optimized via cosine-stress minimization and SGD, yielding semantically meaningful and structurally faithful layouts.}
    \label{fig:embedding}
\end{figure*}

\begin{figure*}[ht]
    \centering
    \includegraphics[width=\textwidth]{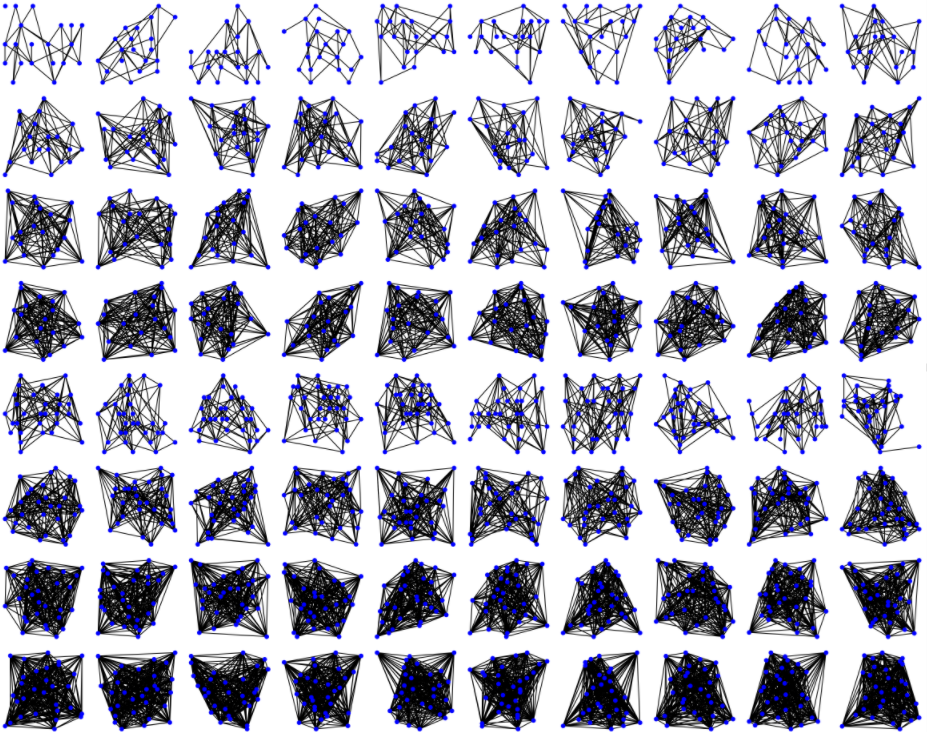}
    \caption{Visualization using the \textsc{Dot} layout. Dot produces layered, hierarchical drawings suited for directed acyclic graphs (DAGs).}
    \label{fig:dot}
\end{figure*}

\begin{figure*}[ht]
    \centering
    \includegraphics[width=\textwidth]{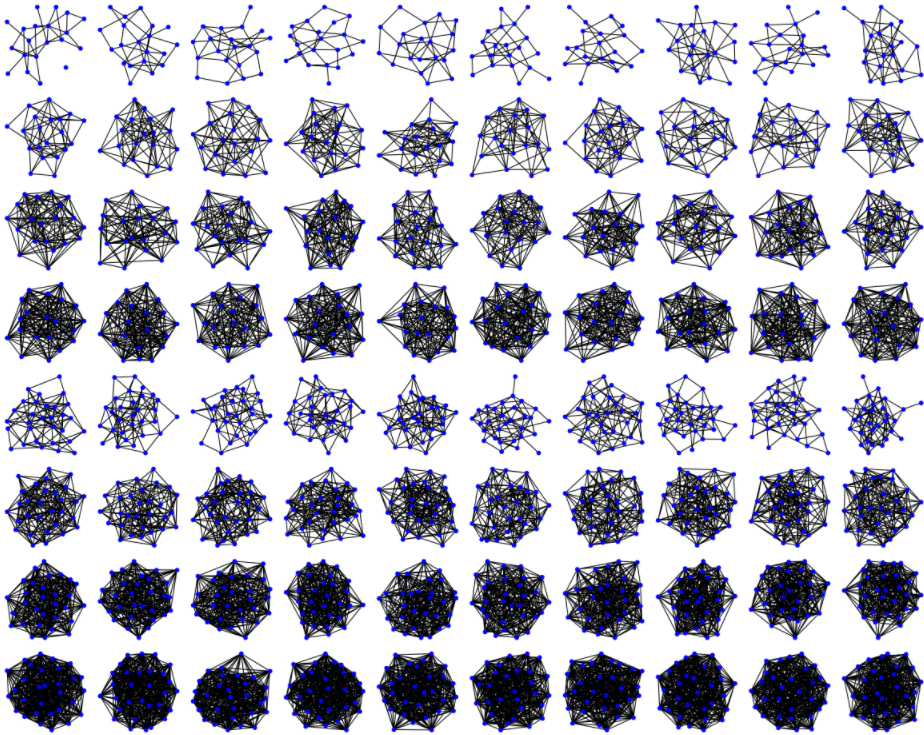}
    \caption{Visualization using the \textsc{Neato} layout. Neato applies a classical force-directed algorithm designed for undirected graphs.}
    \label{fig:neato}
\end{figure*}

\begin{figure*}[ht]
    \centering
    \includegraphics[width=\textwidth]{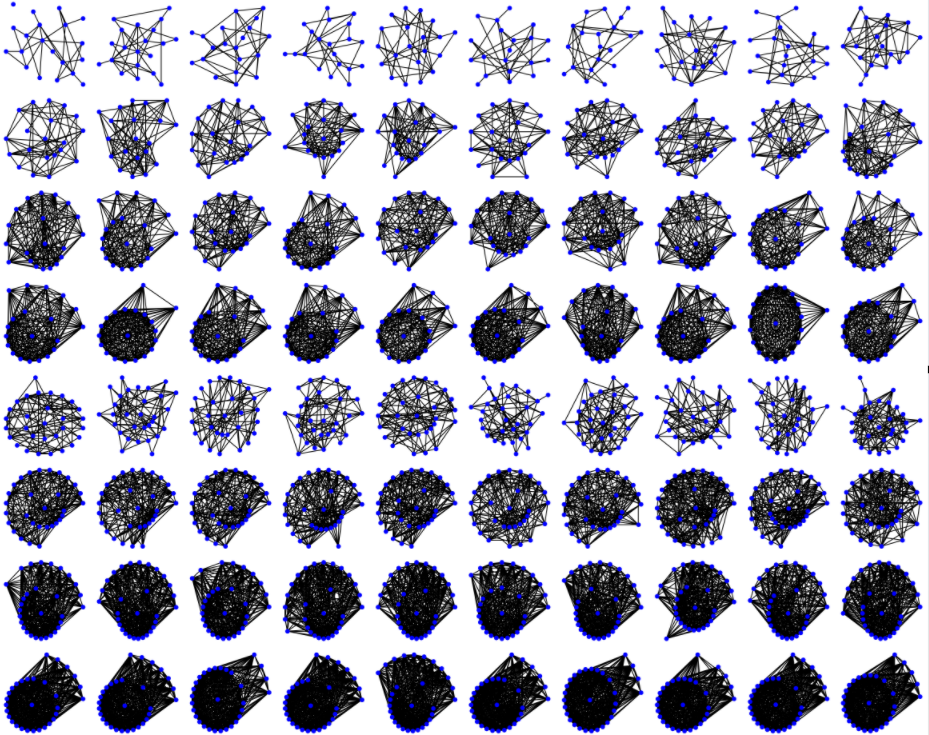}
    \caption{Visualization using the \textsc{Twopi} layout. Twopi arranges nodes in concentric circles, producing radial layouts that emphasize graph depth or centrality.}
    \label{fig:twopi}
\end{figure*}


\begin{figure*}[ht]
    \centering
    \includegraphics[width=\textwidth]{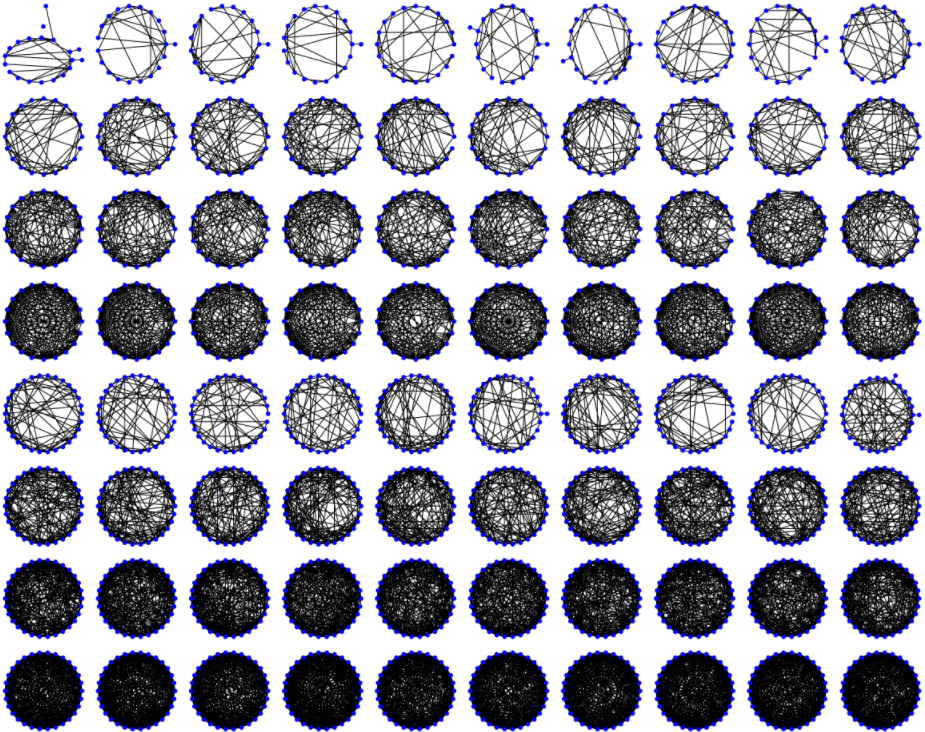}
    \caption{Visualization using the \textsc{Circo} layout. Circo is designed for cyclic graphs, placing nodes in circular formations to highlight loop structures.}
    \label{fig:circo}
\end{figure*}

\begin{figure*}[ht]
    \centering
    \includegraphics[width=\textwidth]{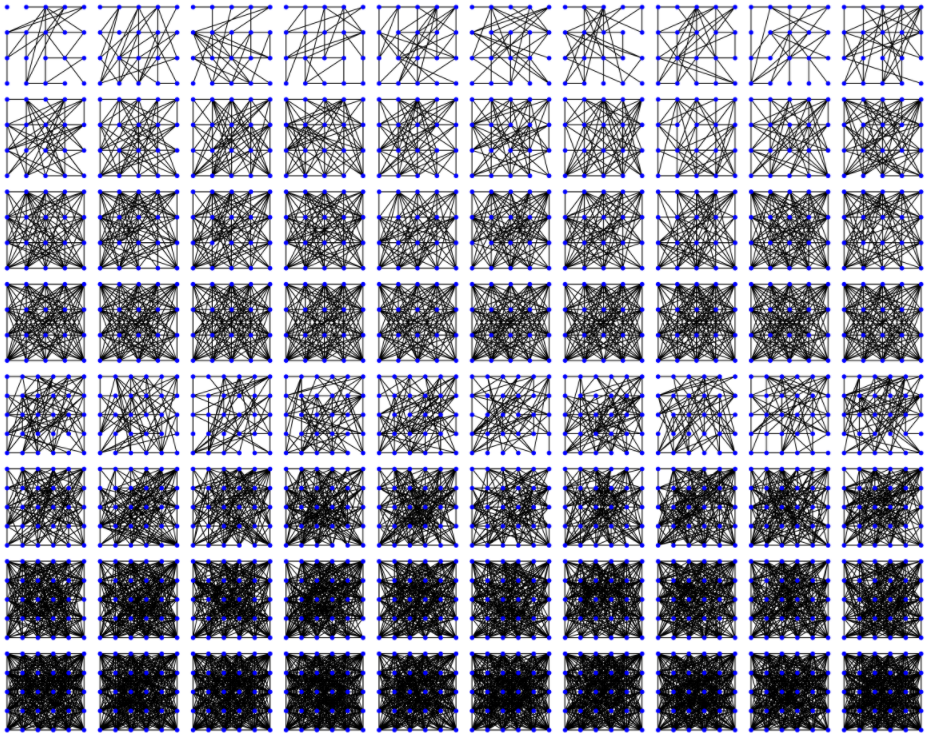}
    \caption{Visualization using the \textsc{Patchwork} layout. Patchwork emphasizes modularity by organizing clusters in distinct spatial regions.}
    \label{fig:patchwork}
\end{figure*}

\begin{figure*}[ht]
    \centering
    \includegraphics[width=\textwidth]{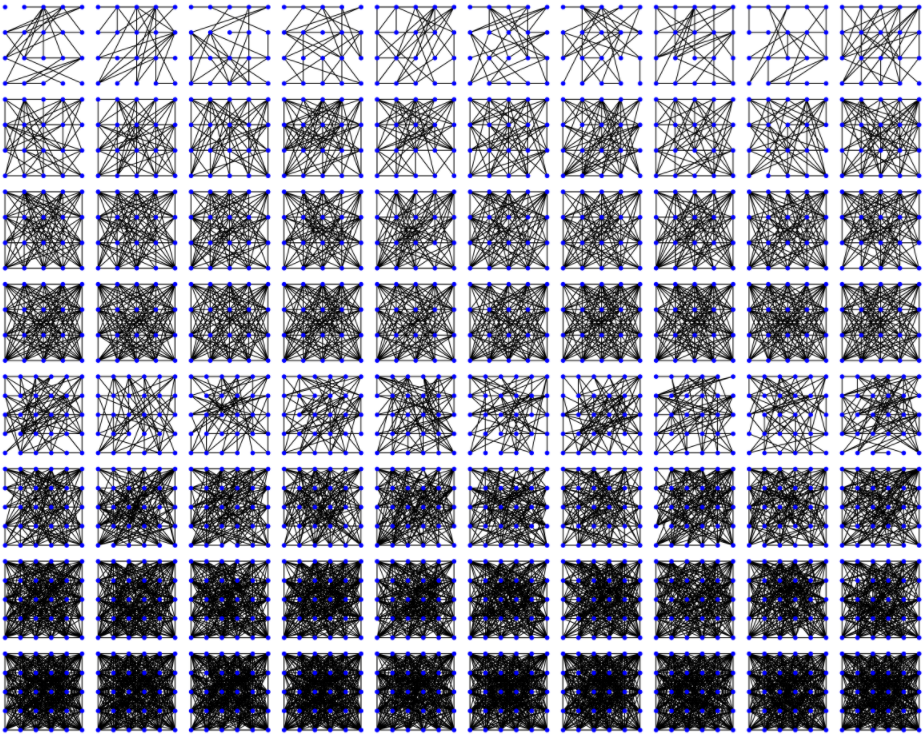}
    \caption{Visualization using the \textsc{Osage} layout. Osage specializes in clustered graphs and maintains modular groupings while minimizing overlap.}
    \label{fig:osage}
\end{figure*}
\end{document}